\documentclass[12pt]{iopart}

\usepackage{iopams}
\usepackage[square,sort&compress,numbers]{natbib}
\usepackage{epsfig}
\usepackage{epstopdf}
\usepackage{color}
\usepackage{url}
\usepackage{bm}
\usepackage{bbm}
\usepackage{amssymb}
\usepackage{eqnarray}
\usepackage{graphicx}

\begin{document}

\title{Relativistic {\bf\em Zitterbewegung} in non-Hermitian 
photonic waveguide systems}

\author{Guanglei Wang}
\address{School of Electrical, Computer, and Energy Engineering,
Arizona State University, Tempe, AZ 85287, USA}

\author{Hongya Xu}
\address{School of Electrical, Computer, and Energy Engineering,
Arizona State University, Tempe, AZ 85287, USA}

\author{Liang Huang}
\address{School of Physical Science and Technology, and Key
Laboratory for Magnetism and Magnetic Materials of MOE, Lanzhou University,
Lanzhou, Gansu 730000, China}

\author{Ying-Cheng Lai} 
\address{School of Electrical, Computer, and Energy Engineering,
Arizona State University, Tempe, AZ 85287, USA}
\address{Department of Physics, Arizona State University, Tempe,
AZ 85287, USA}
\ead{Ying-Cheng.Lai@asu.edu}

\date{\today}

\begin{abstract}

{\it Zitterbewegung} (ZB) is a phenomenon in relativistic
quantum systems where the electron wave packet exhibits a trembling or
oscillating behavior during its motion, caused by its interaction or
coupling with the negative energy state. To directly observe ZB in
electronic systems is difficult, due to the challenges associated with
the atomic scale wavelength of the electron. Photonic systems offer an
alternative paradigm. We exploit the concept of pseudo parity-time
(pseudo $\mathcal{PT}$) symmetry to study ZB in non-Hermitian quantum
systems implemented as an experimentally feasible optical waveguide
array. In particular, the non-Hermitian Hamiltonian is realized through 
evanescent coupling among the waveguides to form a one-dimensional lattice 
with periodic modulations in gain and loss along the guiding direction. As
the modulation frequency is changed, we obtain a number of
phenomena including periodically suppressed ZB trembling, spatial
energy localization, and Hermitian-like ZB oscillations. We calculate
phase diagrams indicating the emergence of different types of dynamical
behaviors of the relativistic non-Hermitian quantum system in an
experimentally justified parameter space. We provide numerical results
and a physical analysis to explain the distinct dynamical
behaviors revealed by the phase diagrams. Our findings provide a
deeper understanding of both the relativistic ZB phenomenon and
non-Hermitian pseudo-$\mathcal{PT}$ systems, with potential applications
in controlling/harnessing light propagation in waveguide-based
optical systems.

\end{abstract}
\noindent{\it Keywords}: pseudo-$\mathcal{PT}$ symmetry, photonic waveguide array, non-Hermitian systems, ZB oscillation, Dirac equation

\maketitle

\section{Introduction} \label{sec:intro}

There has been a great deal of recent interest in investigating the role of
parity-time reversal ($\mathcal{PT}$) symmetry in wave
propagation at different scales with the discoveries of phenomena such as
non-reciprocal beam propagation~\cite{KLB:2005,Rueter:2010,Bender:2013,Peng:2014,
Chang:2014}, and uni-directionally transparent
invisibility~\cite{Ramezani:2010,Feng:2013,Zhu:2014}. The phenomenon
of topologically protected states~\cite{HK:2010,QZ:2011} was originally
discovered in condensed matter physics associated with electronic transport,
but recently it has been demonstrated in optics~\cite{Wang:2008,Wang:2009,
Hafezi:2011,Fang:2012,Rechtsman:2013,Khanikaev:2013,Hafezi:2013}, due
naturally to the correspondence between matter and optical waves. For
example, topological photonics/acoustics were demonstrated by exploiting
the analogy between electronic and synthetic photonic crystals,
where tunable and topologically protected excitations were
observed~\cite{Lu:2014,Yang:2015}. There were also efforts in exploring
new wave features in artificial photonic crystals with/without the
$\mathcal{PT}$-symmetry~\cite{Makris:2008,Longhi:2009,He:2011,Regensburger:2012,
Wimmer:2015,Longhi:2015,Poli:2015,Zhen:2015}. All these have led to the
emergence of a forefront area of research in optics: light propagation in
non-Hermitian $\mathcal{PT}$-symmetric media with balanced loss and gain
profiles. The new field offers the possibility to engineer light propagation,
potentially revolutionizing optics with unconventional applications.

In conventional quantum mechanics, the observable operators are required to
be Hermitian to ensure real eigenvalues. This requirement is the result of
one of the fundamental postulates in quantum mechanics: the physically
observable or measurable quantities are the eigenvalues of the
corresponding operators. However, the seminal works of
Bender {\it et al.}~\cite{BB:1998,BBJ:2002,B:2007} demonstrated that
non-Hermitian Hamiltonians are also capable of generating a purely real
eigenvalue spectrum and therefore are physically meaningful, if the
underlying system possesses a $\mathcal{PT}$ symmetry. This
opened a new research field called non-Hermitian $\mathcal{PT}$ symmetric
quantum mechanics. In fact, in physical systems, non-Hermitian Hamiltonians 
exist in contexts such as electronic transport (in open Hamiltonian systems) 
and gain/loss materials in optics~\cite{moiseyev2011}. Especially, in optics, 
materials with a complex index of refraction can effectively 
be a non-Hermitian $\mathcal{PT}$ symmetric system~\cite{Ruter2010}. 
However, in such a case, the
$\mathcal{PT}$ symmetry constrictions require that the real and the
imaginary parts of the refractive index be an even and odd function in
space, respectively~\cite{RDM:2005,MECM:2008,MAKKC:2012}, which may be 
challenging to be fabricated for experimental study. An alternative
configuration that does not require even/odd spatial
functions is the pseudo-$\mathcal{PT}$ symmetric system with periodic
modulations~\cite{LHZQXKL:2013}. While such a system does not conserve
the energy at any given instant of time because of gain/loss, the energy
does not diverge within any practically long time.

An outstanding problem is how {\em relativistic} quantum effects manifest
themselves in non-Hermitian photonic systems with full or pseudo
$\mathcal{PT}$ symmetry. This is motivated by the tremendous development 
of 2D Dirac materials in the past decade such as 
graphene~\cite{Novoselovetal:2004,Bergeretal:2004,Novoselovetal:2005,
ZTSK:2005,Netoetal:2009,Peres:2010,SAHR:2011}, topological
insulators~\cite{HK:2010,QZ:2011}, molybdenum 
disulfide (MoS$_2$)~\cite{RRBGK:2011,
WKKCS:2012}, HITP [Ni$_3$(HITP)$_2$]~\cite{Sheberlaetal:2014}, and
topological Dirac semimetals~\cite{Liuetal1:2014,Liuetal2:2014}, where
the underlying quantum physics is governed by the relativistic Dirac
equation. It is thus of interest to investigate experimentally more
realizable optical systems hosting relativistic excitations and to study
the manifestations of the fundamental phenomena~\cite{Strange:book} such
as the Klein tunneling, {\em Zitterbewegung} (ZB), and pseudo spin to
exploit their topological origin. In this regard, including gain/loss in 
synthetic optical systems can generate non-Hermitian relativistic 
$\mathcal{PT}$-symmetric excitations through engineering the gain and 
loss in a balanced manner. For example, arranging $\mathcal{PT}$-symmetric 
couplers periodically provides a unique platform to
realize the analogy of the non-Hermitian relativistic quantum systems in
optics~\cite{Longhi:2010,Szameit:2011,SH:2013,Liang:2013}.

In this paper, we focus on a fundamental phenomenon in relativistic
quantum mechanics - ZB oscillations. While the optical analog of the
relativistic ZB effect in Hermitian quantum systems was previously
observed~\cite{DHKTNLS:2010}, we investigate the manifestations of
ZB in non-Hermitian photonic systems. In particular, we consider a binary
waveguide array with periodically modulated imaginary refraction index
in the guiding direction.
This kind of time-periodic gain and loss photonic system is widely used in many
theoretical and experimental
works~\cite{LHZQXKL:2013,JMDP:2014,GW:2015,LJ:2015,LHLMJL:2016, L:2016}.
The optical
waveguide array system, due to its controllable degrees of freedom, has
been a paradigm to study a host of fundamental physical
phenomena~\cite{AW:2012} such as the neutrino oscillations~\cite{MLB:2014},
Bloch oscillation~\cite{PDEBL:1999}, Zener tunneling~\cite{GOGPTW:2005,
TPLMSBP:2006}, and Klein tunneling~\cite{L:2010}. We configure
the system to have a pseudo-$\mathcal{PT}$ symmetry, so that it exhibits
quasi-stationary light propagation with slowly varying time-averaged total
intensity~\cite{LHZQXKL:2013}. By calculating a detailed phase diagram
in an experimentally meaningful parameter space, we uncover a number of
phenomena in the system. In particular, we find that, in certain parameter
regime, the wave amplitude tends to diverge. A remarkable phenomenon is that
there are parameter regimes in which the dependence of ZB oscillations on the
modulation frequency is non-monotonic, where the system exhibits a striking,
periodically suppressed ZB effect with revival or intermittent ZB oscillations 
for low frequencies and a Hermitian-like ZB effect in the high frequency regime.
In the intermediate frequency regime, a surprising spatial energy localization
behavior emerges. By solving the corresponding time-dependent Dirac equation,
we obtain analytic results that provide explanations for the numerically
observed, ZB manifested phenomena. The findings have implications. For
example, the wave divergence phenomenon may be exploited for applications
in optical amplifier and lasing. Intermittent BZ oscillations
can potentially lead to a new mechanism to manipulate/control light
propagation.

Before describing our results in detail, we remark on the two unique
aspects of our work.

First, {\em what results can be considered as constituting a highly 
nontrivial or unexpected general fundamental insight into the ZB phenomenon 
for researchers working in this field?}
Photonic crystals allow researchers to build up classical simulators
of quantum systems. For example, through engineering the gain/loss media,
non-Hermitian, PT-symmetric and non-relativistic quantum physics has been
analogously realized using two sets of evanescently coupled
waveguides~\cite{Ruter2010} through non-reciprocal beam dynamics. Most
previous studies on the optical 
analogies of non-Hermitian PT-symmetric quantum systems are nonrelativistic. 
Quite recently, optical analogies of the Hermitian relativistic Dirac 
equation have been articulated, such as synthetic photonic graphene 
and photonic topological insulators. An outstanding issue concerns the 
PT-symmetry and the related physical effects in synthetic relativistic 
quantum systems. Existing theoretical proposals~\cite{Longhi:2010,SH:2013} 
rely on an exact design of the gain-loss profile or a sophisticated 
strain-based control scheme for the given gain-loss configuration, leading
to stationary, non-Hermitian PT-symmetric systems. Distinct from 
the existing works, our work reports a {\em dynamical} scheme to optically 
realize time-dependent, pseudo-PT symmetric relativistic Dirac equation, and 
we find the striking phenomenon of periodically suppressed and revival ZB 
oscillations. This ``intermittent'' ZB phenomenon is in sharp contrast 
to conventional ZB oscillations with a constant amplitude.
Another aspect of our work lies in observing a resonance boundary between the
pseudo-PT and PT-breaking regions. This kind of resonance boundary has been
reported in previous works, between the conventional PT-symmetric region and
PT-broken region~\cite{JMDP:2014,GW:2015,LJ:2015,LHLMJL:2016,
L:2016}. Here, our results show that the boundary between the pseudo-PT
region and PT-breaking region follows the similar behavior.
   
Second, {\em are there potentially feasible experimental schemes?}
The answer is affirmative.
Experimentally, conventional non-Hermitian PT-symmetry systems 
require that the real part of the effective potential to be an exactly 
even while the imaginary part be an exactly odd function in space.
Our system design {\em relaxes} these requirements and significantly expands
the configuration range of the non-Hermitian systems through introducing 
spatial modulation. The width of ZB trembling is on the order of 
micrometer and the length of the waveguides is millimeters. Concretely, 
our system can be realized in a waveguide array configuration similar to 
that in Ref.~\cite{DHKTNLS:2010}, where the gain effect can be realized, 
for example, through doping of dye molecules such as Rhodamine B 
(emission at 627 nm, excitation at 554 nm - green laser). Loss can be 
introduced using metals evaporated into the waveguides during the 
fabrication process.

In Sec.~\ref{sec:model}, we provide a detailed description of our photonic
waveguide array system with pseudo $\mathcal{PT}$ symmetry, and derive
the underlying Dirac equation. In Sec.~\ref{sec:result}, we present our
main results: phase diagrams, amplitude divergence, regular and revival
ZB oscillations, and spatial energy localization. In Sec.~\ref{sec:conclusion},
we summarize the main results and offer a discussion. Certain details of 
the analytical derivation based on the relativistic quantum 
Dirac equation are given in Appendix.

\section{Optical waveguide system with pseudo $\mathcal{PT}$ symmetry and
description in terms of the Dirac equation} \label{sec:model}

\begin{figure}
\centering
\includegraphics[width=\linewidth]{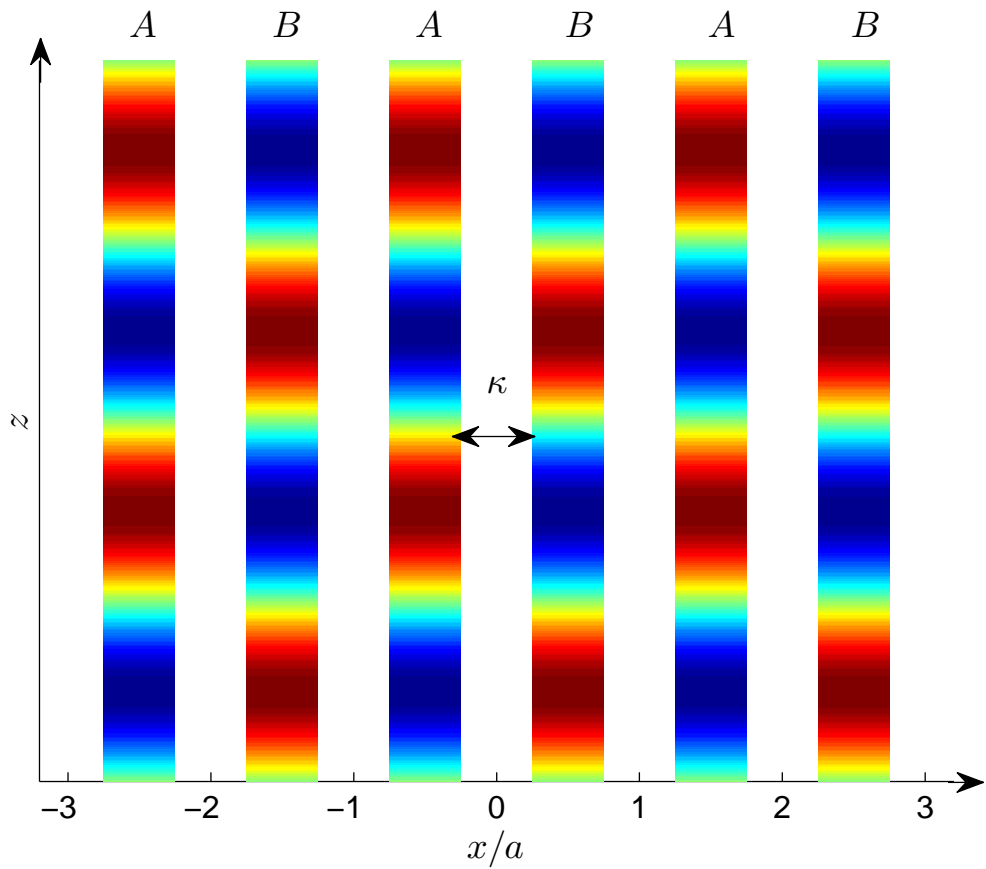}
\caption{ {\bf Schematic illustration of the system configuration}.
Two sets of waveguides, marked as $A$ and $B$, are interleaved with each other.
The real parts of the refractive index for $A$ and $B$ are different, and their
imaginary parts are small but have opposite signs. The distance between two
adjacent waveguides is $a$ and they are coupled through constant strength
$\kappa$. The imaginary part of the refractive index changes along the $z$
direction, mimicking a time dependent non-Hermitian potential.}
\label{fig:schematic}
\end{figure}

We consider a two-dimensional binary photonic superlattice consisting of
two kinds of interleaved waveguides, $A$ and $B$, as shown in
Fig.~\ref{fig:schematic}. The distance between two adjacent waveguides is
$a$ and the coupling strength between them in the effective tight-binding 
Hamiltonian is $\kappa$. Due to the similarity between the optical and 
electronic waves, an optical waveguide system can simulate the time evolution
of the electronic wavefunction subject to an equivalent potential. If the 
real parts of the refractive index of the nearest neighboring waveguides 
have the same mismatch, the waveguide superlattice can simulate the behaviors 
of the finite-mass Dirac equation. In particular, the dispersion relation 
of the lattice is a hyperbola about the edge of the first Brillouin 
zone~\cite{DHKTNLS:2010,L:2010}, and the $A$ and $B$ sublattices can be 
regarded as corresponding to the two components of the spinor wavefunction
underlying the Dirac equation. When the incident angle of a wave packet is
approximately the first Bragg angle, for an initial group velocity close 
to that of the edge of the first Brillouin zone, the optical system
obeys the relativistic Dirac equation. Experimentally, it is difficult to
realize a waveguide system with time-dependent refractive index. 
However, the $z$ direction represents effectively the time dimension. The
two-dimensional binary photonic superlattice system is then completely
equivalent to a time dependent, one-dimensional relativistic Dirac system,
making possible experimental studies of various relativistic quantum effects.

We employ the tight-binding approximation to analyze the waveguide 
superlattice. The amplitude of the optical field is described by
\begin{equation} \label{eq:wave}
i\frac{da_n}{dz}=-\kappa(a_{n+1}+a_{n-1})+(-1)^n\sigma(z)a_n,
\end{equation}
where $a_n$ is the amplitude of the optical field in the $n$th waveguide, 
$\kappa$ is the effective coupling constant, and $\sigma$ is the propagation 
mismatch. We assume that the real part of the mismatch $\sigma_r$ between 
the nearest neighboring waveguides is constant. The
space dependence of $\sigma(z)$ is embedded in its imaginary part:
$\sigma_i(z)=r\sigma_r\sin{(\omega z)}$, where $\sigma_r$ and $\sigma_i$ 
denote its real and imaginary parts, respectively, with $r$ being the ratio
between them that characterizes the strength of gain/loss.
The quantity $\omega$ is the frequency of spatial gain/loss modulation 
in the $z$-direction. Equivalently, $\omega$ can be regarded as the
frequency of the imaginary part of the time-varying potential for 
the corresponding quantum system that is fundamentally non-Hermitian
due to the imaginary potential. We note that, for 
a Hermitian system, i.e., a superlattice system with a purely real 
refractive index, previous works~\cite{DHKTNLS:2010,L:2010} showed the
emergence of two minibands with the dispersion relation given by
$\omega_{\pm}(q)=\pm\sqrt{\sigma^2+4\kappa^2\cos^2{(qa)}}$. For the 
wavevector along the $x$-direction, $q$ is close to $\pi/(2a)$ and the 
quantities $\omega_{\pm}(q)$ exhibit a hyperbola-like behavior with a 
gap of $2\sigma$, which is similar to the electron and positron 
dispersion curves from the finite-mass Dirac equation.
For a general non-Hermitian system, it is straightforward to show that 
the dispersion relation has the same form as that for a Hermitian system
(see Appendix), but there are two differences: (a) the quantity $\sigma$ 
becomes complex and (b) the system is time dependent due to the external 
driving. In the limit of small imaginary part, the complex frequency in 
the dispersion relation possesses the following real part 
\begin{displaymath}
    \pm\sqrt{(\sigma_r^2-\sigma_i^2(z))+4\kappa^2\cos^2{(qa)}}, 
\end{displaymath}
guaranteeing that our non-Hermitian system can still retain the equivalence 
to the relativistic quantum system for a massive particle at the boundaries
of the Brillouin zone. This condition constricts the incident angle
of the wave packet to be about $\theta_B\approx\lambda/(4n_sa)$, where 
$\lambda$ is the wavelength and $n_s$ is the refractive index of the substrate. 
We emphasize that this analogy holds only when the wave packet 
is located near the Brillouin zone boundary. Analogy in other regions would
be affected by the dispersive effect of the waveguides. In fact, many 
experiments have shown that the analogy can hold up to hundred
millimeters~\cite{L:2010,L:2011,KZDHTNS:2013}.

To demonstrate the equivalence of the waveguide equation to 
the Dirac equation, we make the following
substitutions~\cite{DHKTNLS:2010,L:2010,KZDHTNS:2013}:
\begin{equation} \label{eq:subs}
[\psi_1,\psi_2]^T=[(-1)^na_{2n},i(-1)^na_{2n-1}].
\end{equation}
From the periodicity and symmetry of the system, we see that a unit cell
from sublattice $A$ combining with one from sublattice $B$ is 
effectively a unit cell that forms a Dirac spinor. It is necessary to use 
the dimensionless continuous transverse coordinate 
$\xi\leftrightarrow n=x/(2a)$ to form a spatial derivative operator. With 
these considerations, we obtain the following equivalent one-dimensional 
Dirac equation
\begin{equation} \label{eq:Dirac}
i\frac{\partial\psi}{\partial z}=-i\kappa\alpha_1
\frac{\partial\psi}{\partial\xi}+\sigma(z)\alpha_3\psi,
\end{equation}
where
\begin{equation} \label{eq:Pauli}
    \alpha_1= \left( \begin{array}{cc}
            0 & 1 \\ 
            1 & 0
                     \end{array} \right)
    \alpha_3= \left( \begin{array}{cc}
            1 & 0 \\ 
            0 & -1
              \end{array} \right),
\end{equation}
are the Pauli matrices. These considerations hold when $\sigma$ is complex.
Comparing with the standard Dirac equation, we note the following equivalence:
\begin{equation} \label{eq:eqi}
\kappa\leftrightarrow c, \sigma\leftrightarrow mc^2/\hbar,
\end{equation}
i.e., the coupling constant between the waveguides corresponds to the
speed of light while the mismatch between the waveguides is equivalent
to the mass of the underlying particle. Indeed, as we can see from the
dispersion relation of the superlattice, the band gap for the Brillouin
zone is also proportional to $\sigma$, corresponding to the dispersion
relation of a massive relativistic quantum particle.

In our simulations, we assume that the initial optical field has the
form $E(x,0)=G(x)e^{2\pi ixn_s\theta_B/\lambda}$, where $G(x)$ is a
Gaussian beam envelope. The superlattice is made of $200$ waveguides
separated from each other with the distance $a=16\mu\mbox{m}$, and
the effective coupling rate is $\kappa=0.14\mbox{mm}^{-1}$. The Gaussian
beam has the wavelength of $\lambda=633\mbox{nm}$ with a spot size about
$105\mu\mbox{m}$. The modulation frequency is scaled by the factor
$\omega_0=1/a$ and the lengths are normalized by $a$.

\section{Results} \label{sec:result}

\subsection{Emergence of pseudo-$\mathcal{PT}$ symmetry}

\begin{figure}
\centering
\includegraphics[width=0.8\linewidth]{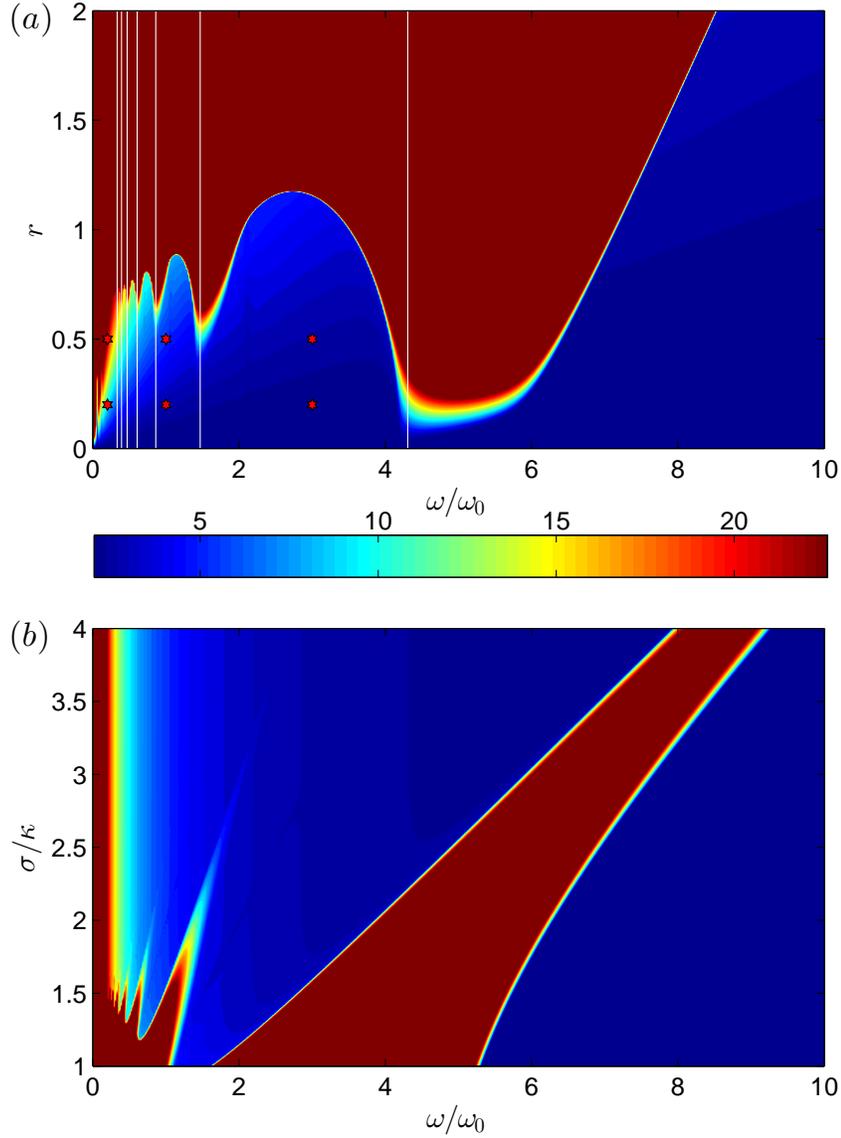}
\caption{{\bf Phase diagrams of relativistic quantum
photonic superlattice system}. The dark red region indicates the 
pseudo-$\mathcal{PT}$ breaking phase, and the other region corresponds to 
the pseudo-$\mathcal{PT}$ phase. (a,b) Phase diagrams in the $r-\omega$
and $\sigma_r-\omega$ parameter plane, respectively. The color is coded
in terms of the logarithm of the maximum intensity in the specific
parameter region. The white lines in (a) denote the positions of the valleys
in the oscillatory variations of the boundary. The first two vertical lines, 
counted from right to left, correspond to the two sets of parameters shown in
Fig.~\ref{fig:oscillation}. The red stars in (a) specify the positions of the
parameters shown in Fig.~\ref{fig:ZB}. For each phase diagram, the
computational grid in the corresponding parameter plane has the size
$512\times512$. The parameters are $\sigma_r/\kappa=2.1$ (a) and 
$\sigma_i/\kappa=0.5$ (b). }
\label{fig:PD}
\end{figure}

To ascertain the emergence of the pseudo-$\mathcal{PT}$ symmetry,
we probe into the parameter space of the photonic superlattice system
systematically by calculating the phase diagrams. The results are shown
in Fig.~\ref{fig:PD}, where the same initial wave packet with a group
velocity determined by the relativistic Dirac point is used for different
parameter combinations. The evolution is simulated for a relatively
long time to ensure that the system has settled into a steady state.
Figures~\ref{fig:PD}(a) and \ref{fig:PD}(b) show the phase diagrams in
the $r-\omega$ and $\sigma_r-\omega$ parameter planes, respectively. The
color is coded in terms of the maximum value of the intensity of the
optical field for the corresponding parameters. Considering the fact
that, in the pure gain regime, the intensity grows exponentially,
we use a logarithmic function to rescale the intensity for better
visualization. We set a cut-off intensity level beyond which the
optical field is deemed to have diverged, corresponding to the
$\mathcal{PT}$ breaking case. The cut-off criterion we used is
$E_{tot}/E_{tot}^0\sim10^9$, where $E_{tot}$ is the maximum of the total
intensity and $E_{tot}^0$ is the total intensity when there is no modulation.
Even for this relatively high cut-off intensity, there is little change in 
the shape of the boundary. In most parts of the boundary of the 
$\mathcal{PT}$ breaking phase and the pseudo-$\mathcal{PT}$ phase, the 
change in the color from red to blue are quite rapid.
In Figs.~\ref{fig:PD}(a) and \ref{fig:PD}(b),
the divergent regions are coded as dark red. For a small modulation frequency,
a number of waveguides absorb energy within the simulation time so that
their intensity can exceed the cut-off value. For such waveguides,
a pseudo-$\mathcal{PT}$ behavior cannot be numerically detected.
Because of this, there is a persistently red region in the phase diagrams
when the modulation frequency is close to zero. From Fig.~\ref{fig:PD}(a),
we see that a pseudo-$\mathcal{PT}$ behavior exists either in the parameter
region with small gain/loss strength or in the region with a high modulation
frequency. The reason is that, when the gain/loss of the waveguides is small,
energy absorption is insignificant so that the waveguides will have
sufficient time to transfer the energy within the superlattice and dissipate
the absorbed energy into the adjacent waveguides. When the modulation
frequency is high, the period becomes small so that the waveguides are
able to dissipate the absorbed energy through itself when the imaginary
part of the refractive index changes its sign during light propagation.
In general, the stronger the gain/loss strength, the higher the
modulation frequency is needed to balance the gain and loss, providing an
explanation for the shape of the boundary of the two regions in the high
modulation frequency regime in the phase diagrams. Figure~\ref{fig:PD}(b)
exhibits a similar behavior, i.e., the pseudo-$\mathcal{PT}$ behavior exists
in the larger real refractive index (corresponding to weaker gain and loss
effects) and higher modulation frequency regions. A general feature is that
higher modulation frequency is favorable for the emergence of the
pseudo-$\mathcal{PT}$ symmetry.

A phenomenon present in both Figs.~\ref{fig:PD}(a,b) is
that the boundary of the pseudo-$\mathcal{PT}$ symmetric and the non
pseudo-$\mathcal{PT}$ regions exhibits oscillatory variations as
the modulation frequency is changed. For $\omega/\omega_0\sim 4-6$, the
system exhibits a divergent behavior even for the small gain/loss strength.
To understand the origin of the
boundary variations, we examine the details of the time evolutions of
the system. In Fig.~\ref{fig:PD}(a), we mark the positions of the valleys
in the variation with white lines, which are $\omega/\omega_0=0.6, 0.9, 1.5$
and $4.2$, leading to the ratio of the period, i.e., $1/\omega$, to be
about $7:5:3:1$, indicating that the system tends to diverge when the
modulation period is odd times of the rightmost line. When the ratio is
even, the system tends to stay in the pseudo-$\mathcal{PT}$ phase.
    Here we want to emphasis that this kind of resonance boundary has been
    extensively studied in the conventional PT-symmetric
    system~\cite{JMDP:2014,GW:2015,LJ:2015,LHLMJL:2016, L:2016}.
However,
when the ratio becomes large, it is hard to observe the oscillatory behavior, 
due mainly to the finite resolution of the simulation.
Another reason is
that the system tends to shift into the high energy regime when the 
modulation frequency is low, so the oscillations are buried in the regime 
that does not correspond to $\mathcal{PT}$ breaking. The boundary in 
Fig.~\ref{fig:PD}(b) exhibits a similar variational behavior. Particularly, 
as $\sigma_r$ becomes larger, the imaginary part becomes relatively less 
significant. As a result, the band gap and thus the ZB trembling frequency 
is given by $2\sigma_r$ (to be described in Sec.~\ref{subsec:ZB}). When 
the modulation frequency is equal to the ZB trembling frequency, a resonant 
effect appears, forming a pseudo-$\mathcal{PT}$ phase boundary that starts 
from $\omega/\omega_0 = 2$ in Fig.~\ref{fig:PD}(b). The specific ratio 
relation suggests the occurrence of resonances in the system that can 
enhance or suppress the gain and loss of the superlattice. A more detailed 
understanding can be obtained through the dynamical oscillations in the 
system, i.e., the ZB effect.  

\subsection{Relativistic {\bf\em Zitterbewegung} in photonic superlattice}
\label{subsec:ZB}

\begin{figure}
\centering
\includegraphics[width=0.8\linewidth]{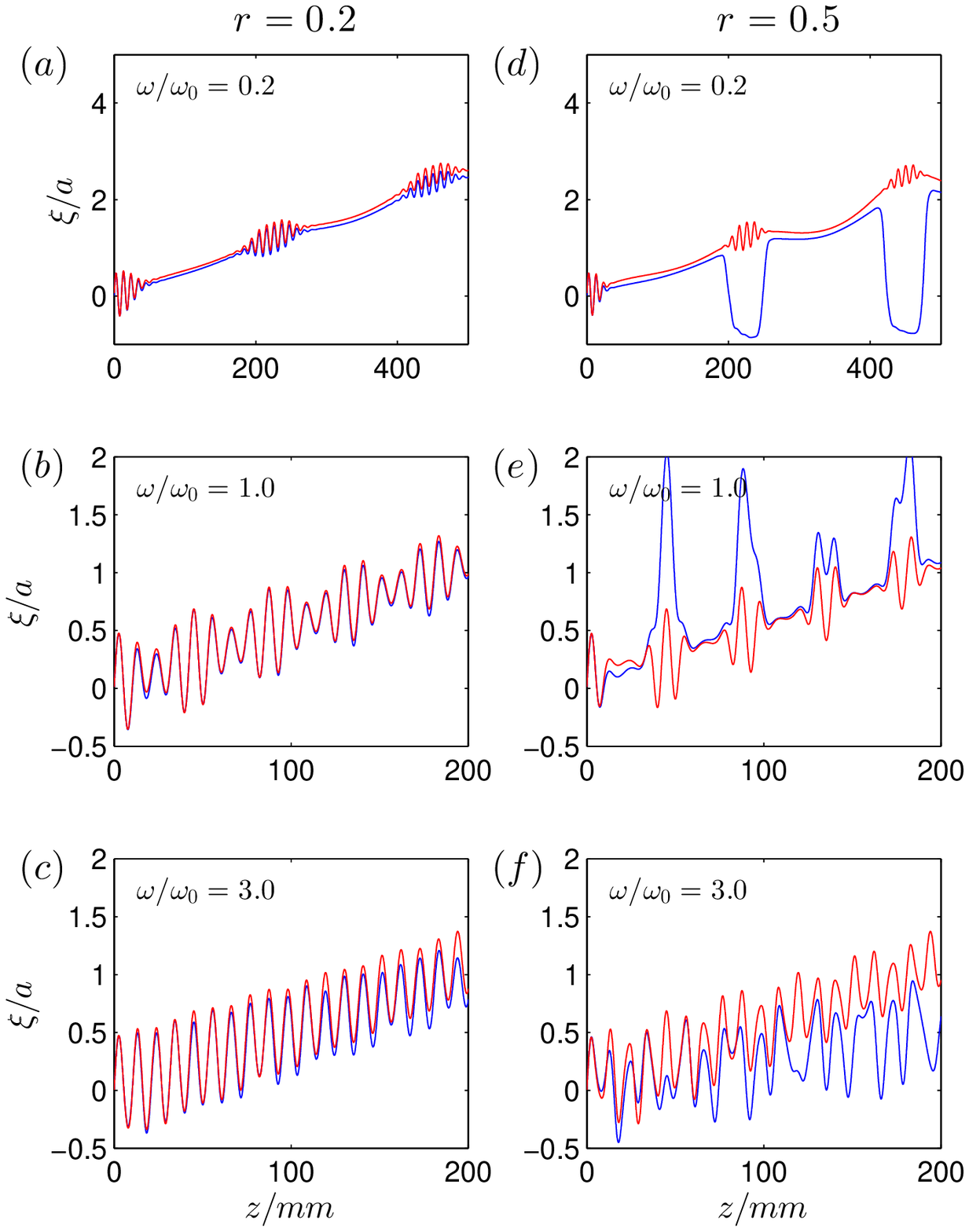}
\caption{{\bf Relativistic {\bf\em Zitterbewegung} in 
non-Hermitian photonic superlattice systems}. 
The six panels correspond to all possible combinations of $r=0.2, 0.5$ 
and $\omega/\omega_0=0.2, 1.0, 3.0$, as marked by the red stars in 
Fig.~\ref{fig:PD}(a). (a-c) Mean/expectation values of the position of 
the wave packet for $r=0.2$ and $\omega/\omega_0=0.2, 1.0, 3.0$,
respectively. (d-f) Similar plots but for $r=0.5$. The blue and red 
curves correspond to the beam center of mass of the wave packet from the
simulation results and the mean expectation value of the position 
operator from the analytic results, respectively.}
\label{fig:ZB}
\end{figure}

ZB is a purely relativistic quantum effect resulting from the interference
between the positive and negative energy states of the Dirac fermion. To
experimentally observe ZB in electronic systems is challenging due to
the small wavelength and the extremely high oscillating frequency.
Photonic superlattice systems with the underlying equation having the same 
mathematical form as the Dirac equation provide an alternative paradigm for
detecting and characterizing the ZB phenomenon~\cite{DHKTNLS:2010}.
Our goal is to investigate whether ZB can emerge in pseudo-$\mathcal{PT}$
symmetric photonic systems. Figure~\ref{fig:ZB} shows a number of
representative time series of the beam center of mass of the wave
packet for different parameters. There are apparent oscillations
in the time series. Since oscillations are a generic feature of the wave
equations, the issue is whether these oscillations are true manifestations
of ZB. We address this issue by analyzing the equivalent Dirac equation
to determine if the oscillations have a relativistic quantum origin.
The basic idea is to write the Dirac equation in the momentum space 
to obtain the time evolution of the wavefunction~\cite{L1:2010}. In order
to arrive at an analytical form, we make the assumption that the total
Hamiltonian at any two given instants of times are commutative so that
a sophisticated treatment of the time ordering operator is
not necessary. Note that, while for time independent quantum systems time 
ordering is not necessary, for a time dependent Hamiltonian this may not 
be the case. Nonetheless, we expect that, in the parameter regime where 
the non-Hermitian effect is weak, i.e., small $\sigma_i$, our analytic
results would agree with the direct simulations results.

The main results of our analysis can be summarized by dividing the time 
evolution of the expectation value of the position operator into three 
components: the drift $\xi_d(t)$, the ZB trembling component $\xi_{ZB}(t)$, 
and the purely imaginary component $\xi_{Im}(t)$. We have (detailed
derivation can be found in Appendix)
\begin{eqnarray} \label{eq:ana}
    \nonumber
    &\langle\xi\rangle(t)&=[\xi_{d}(t)+\xi_{ZB}(t)+\xi_{Im}(t)]/|\psi(t)|^2,
    \\
    &\xi_d(t)&=4\pi\int_{-\infty}^\infty
    dkG^2(k)\kappa^3k^2\frac{|\cos(A)|^2A^*+|\sin(A)|^2A}{A|A|^2}t^3, \\
    \nonumber
    &\xi_{ZB}(t)&=4\pi\int_{-\infty}^\infty
    dkG^2(k)\sin(A)\cos^*(A)\frac{\kappa (-\sigma_rt+i\sigma_i(\cos(\omega
    t)-1)/\omega)^2}{A^3}t, \\ \nonumber
    &\xi_{Im}(t)&=8\pi i\int_{-\infty}^\infty dkG(k)\partial_kG(k)\kappa
    k \mbox{Im}(\frac{\sin(A)\cos^*(A)}{A})t,
\end{eqnarray}
where 
\begin{displaymath}
A(t)=\sqrt{\kappa^2k^2t^2+(\sigma_rt-i\sigma_i
\frac{\cos(\omega t)-1}{\omega})^2} 
\end{displaymath}
is the normalization constant associated with the
Euler formula for the Pauli matrices, and
\begin{displaymath}
G(k)=(1/2\pi)\int_{-\infty}^\infty d\xi G(2d\xi)e^{-ik\xi}
\end{displaymath}
is the angular spectrum of the Gaussian beam envelop. The modulus of the 
wavefunction is 
\begin{eqnarray} \label{eq:psi_t}
|\psi(t)|^2 & = & 4\pi\int_{-\infty}^\infty dkG(k)^2[|\cos(A)|^2 \\ \nonumber
& + & |\frac{\sin(A)}{A}|^2(\kappa^2k^2t^2+\sigma_r^2t^2+(\frac{\cos(\omega t)-1}{\omega}\sigma_i)^2)],
\end{eqnarray}

We note that, for a relativistic quantum system described by the Dirac
equation, the time evolution of the expectation value of the position
operator contains the drift and the ZB trembling components only. For
a stationary Hermitian system, we have $A=A^*\propto t$, the drift
component can be simplified as 
\begin{displaymath}
\xi_d(t)\propto \frac{A\cos^2(A)+A\sin^2(A)}{A^3}t^3\propto t,
\end{displaymath}
which describes the motion of the wave packet at constant velocity. The ZB 
trembling component $\xi_{ZB}(t)$ can be simplified as an oscillatory term 
described by $\sin(A)\cos(A) \propto \sin(2A)$, and the purely imaginary 
component $\xi_{Im}(t)$ is simply zero.

\begin{figure}
\centering
\includegraphics[width=0.8\linewidth]{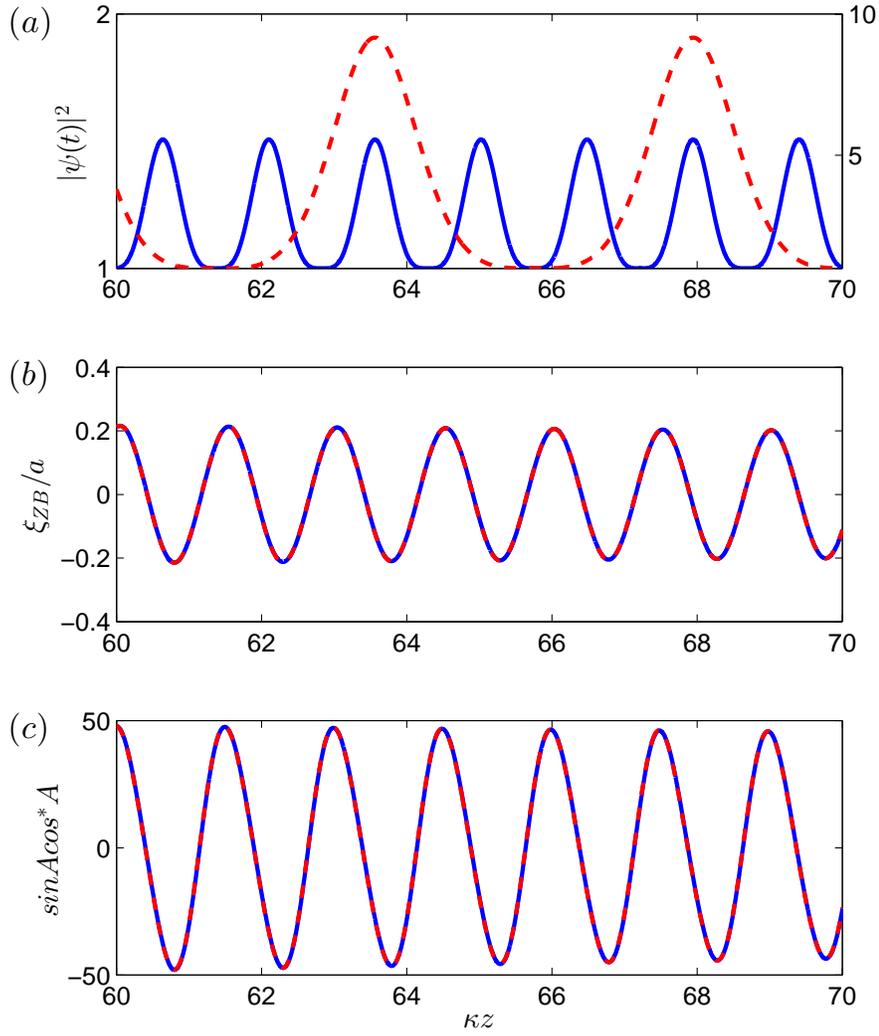}
\caption{{\bf Oscillatory behaviors in the wave packet}. 
Time series (equivalently, propagation along the $z$ direction) of the
photonic superlattice system for the parameters corresponding to the
valleys of the boundary variations in the phase diagram (Fig.~\ref{fig:PD}).
(a-c) Time series of the modulus of the Dirac spinor of the two sublattices,
of ZB trembling, and of the oscillation term $\sin(A)\cos^*(A)$ in $\xi_{ZB}$,
respectively. The parameter setting is $\sigma_r/\kappa=2.1$, $r=0.5$, 
$\omega/\omega_0=4.2$ (blue solid curves), and $\omega/\omega_0=1.5$ (red 
dashed curves).}
\label{fig:oscillation}
\end{figure}

The analytical results can be used to interpret the numerically observed
dynamical behaviors of the system associated with the boundary
variations in the phase diagrams in Fig.~\ref{fig:PD}. Specifically,
Fig.~\ref{fig:oscillation} shows the analytical results for $r=0.5$
and $\omega/\omega_0=4.2, 1.5$ (the blue solid and red dashed curves, 
respectively), which correspond to the two rightmost white lines marked 
in Fig.~\ref{fig:PD}(a). Figure~\ref{fig:oscillation}(a)
shows the time evolution of the normalized modulus of the two sublattices
$A$ and $B$. As expected, the modulus of the red dashed curve is about three
times slower than that of the blue solid curve, which is the ratio between
the modulation frequencies. However, the time
evolution of the ZB trembling term $\xi_{ZB}(t)$ has the same frequency
for the two sublattices, as shown in Fig.~\ref{fig:oscillation}(b),
indicating that ZB trembling has little dependence on the modulation
frequency. A comparison of the time series in Figs.~\ref{fig:oscillation}(a,b) 
indicates that the frequency of ZB oscillations is close to that of the 
modulus oscillation for $\omega/\omega_0=4.2$. 

Note that, for $r=0$ the band gap is 
$2\sigma_r=4.2$. These results imply that the variations in the
boundary between the pseudo-$\mathcal{PT}$ symmetric and the
pseudo-$\mathcal{PT}$ symmetry breaking regions are due to the resonant
interaction between the modulation to the waveguides and ZB trembling.
The energy absorption process is enhanced when the ratio between the
modulation period and the ZB trembling period is an odd integer.
For the high frequency region where the modulation period is smaller than 
the ZB trembling period, the system exhibits a relatively simple behavior,
where the boundary can be described by a linear relation between $r$ and
$\omega/\omega_0$.

Figure~\ref{fig:ZB} shows the simulation (blue) and analytic (red) results
for the ZB effect, where the behavior of the real part of the quantity
$(\xi_d(t)+\xi_{ZB}(t))/|\psi(t)|^2$ is illustrated for six different
parameter combinations indicated by the red stars in Fig.~\ref{fig:PD}(a).
When the imaginary part of the refractive index is relatively small, i.e.,
when the system is only weakly non-Hermitian, the analytical and simulation
results agree well with each other, as shown in Figs.~\ref{fig:ZB}(a-c)
for $r = 0.2$. Note that the scale of the $x$ axis in Fig.~\ref{fig:ZB}(a) 
is about three times of that in Fig.~\ref{fig:ZB}(b-c), and
the periods of the oscillations for these cases are the same, indicating
that the oscillation periods have little dependence on the modulation
frequency. Our analytical results provide a base to attribute the
oscillations as resulting from the ZB trembling term $\xi_{ZB}(t)$,
demonstrating their relativistic quantum origin. We note that, for a 
Hermitian system, the dynamical evolution of the wave packet can be
represented as the superposition of a constant drift behavior and simple
sinusoidal oscillations, but the dynamics of ZB oscillations in the
relativistic non-Hermitian system are richer. Specifically, from
Fig.~\ref{fig:ZB}(a), we see that, for low modulation frequencies, 
ZB oscillations can be enhanced or suppressed in different time intervals.
However, for high modulation frequencies [e.g., Fig.~\ref{fig:ZB}(c)],
ZB oscillations are conventional in the sense that there are
no apparent enhancement or reduction effects. An intermediate situation
arises between these two cases, e.g., for $\omega/\omega_0=1.0$.
These results indicate that the modulation frequency of the refractive
index can affect the amplitude of ZB oscillations.

The role of the modulation frequency in the amplitude evolution of ZB
oscillations can be assessed in a more detailed manner through a close
examination of the oscillation term in $\xi_{ZB}(t)$, i.e., $\sin{A}\cos^*{A}$,
for different values of the modulation frequency, as shown in 
Figs.~\ref{fig:oscillation}(b-c), where the oscillating patterns have the
same modulation frequency, implying that the frequency of ZB
oscillations is entirely determined by its sine and cosine components.
Further, the ZB oscillations for the two different values of the modulation
frequency are in pace with each other in time, indicating that the modulation
frequency has little effect on the frequency of ZB oscillations. This
is expected because the quantity $A(t)$ is an integration of $\sigma(t)$ so
that the amplitude of $\sigma(t)$ is a key factor. Nonetheless, the 
modulation frequency does affect the oscillating intensity,
as shown in Fig.~\ref{fig:oscillation}(a). Thus, the envelope behavior of
ZB oscillations in Figs.~\ref{fig:ZB}(a-c) is determined by the
modulation of the periodic refractive index: for high and low modulation
frequencies, ZB oscillations are suppressed and enhanced, respectively.
A conclusion is that, for high modulation frequency and low modulation
intensity, the effect of refractive-index modulation is weak, as exemplified
in Fig.~\ref{fig:ZB}$(c)$. For $r = 0.5$, the analytically predicted 
amplitude of the oscillation behavior [Figs.~\ref{fig:ZB}(d-f)] 
deviates from the simulation results (except for the initial phase of 
the dynamical evolution). In spite of the disagreement, the predicted 
phase behavior of the oscillations agrees with
the simulation results. The failure of the analytic theory to predict
correctly the amplitude behavior of ZB oscillations stems from the
hypothesis used in our analysis: the Hamiltonians at different times are
commutative. This hypothesis is violated when the non-Hermitian effect
becomes pronounced, i.e., as the imaginary component of the refractive
index and the time dependent modulation are relatively large. Indeed,
as we increase the value of $r$ from $0.2$ to $0.5$, the agreement becomes
increasingly worse, especially for
large time. However, for the parameter region on the right side of
the phase diagrams in Fig.~\ref{fig:PD} where the imaginary part of the
refractive index is small, a reasonably good agreement between the analytical
and simulation results is obtained, due to the relatively weak non-Hermitian
effect. For the high modulation frequency region (the rightmost part of
Fig.~\ref{fig:PD}), ZB oscillations are similar to those of the
conventional case where the modulation in the refractive index is absent.
Physically, in the high modulation frequency region, the waveguides do not have
time to absorb and dissipate energy, generating a mean-field like effect.

\begin{figure}
\centering
\includegraphics[width=0.8\linewidth]{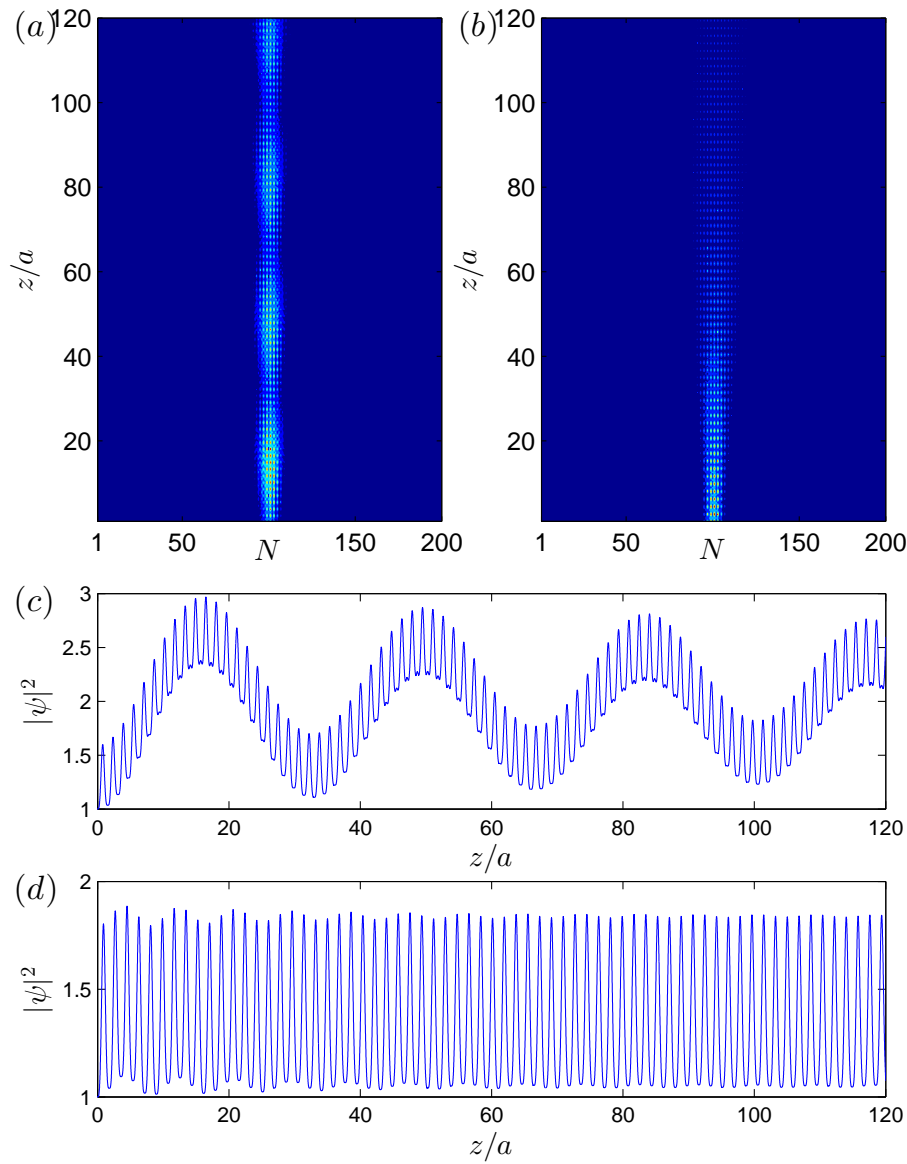}
\caption{{\bf Phenomenon of spatial energy localization in
photonic superlattice}. Spatial view of the field intensity and time
evolution of the modulus of the wave function for $r = 0.5$, (a,c)
$\omega/\omega_0=4.0$, and (b,d) $\omega/\omega_0=3.5$.}
\label{fig:soliton}
\end{figure}

In the intermediate frequency regime, the analytical and simulation results
do not agree with each other. Numerically, we find the phenomenon of spatial
energy localization, as shown in Fig.~\ref{fig:soliton}. In particular, in the
parameter space near the boundary of the phase diagram, e.g., for $r=0.5$ and
$\omega/\omega_0=4.0$, spatially the wave energy is localized within a small
region about the center of the waveguide array, as shown in
Fig.~\ref{fig:soliton}(a). This phenomenon of energy localization is
sensitive to the value of $\omega$. Figure~\ref{fig:soliton}(b) shows a
contrary example for $\omega/\omega_0=3.5$, where the wave packet spreads
after a short time. We find that the wave spread becomes faster as the
value of $\omega$ is reduced. Figures~\ref{fig:soliton}(c,d) show the total
wave intensity corresponding to the cases in Figs.~\ref{fig:soliton}(a,b),
respectively. We observe an apparent envelope modulation behavior in the
case of energy localization. The strong reduction in the spread of the
wave packet is similar to that reported in Ref.~\cite{DVL:2013}, where
a localization behavior in quantum diffraction was found to result from
the linearization of the quasienergy spectrum close to the
$\mathcal {PT}$-symmetry breaking boundary. Generally, this is a dynamical
localization phenomenon in one-dimensional driving
systems~\cite{LMLRLCF:2006,IAWDS:2007,KVKT:2016,DK:1986}.
The width of the localized wave packet depends on the width of the initial
wave packet.

\begin{figure}
\centering
\includegraphics[width=\linewidth]{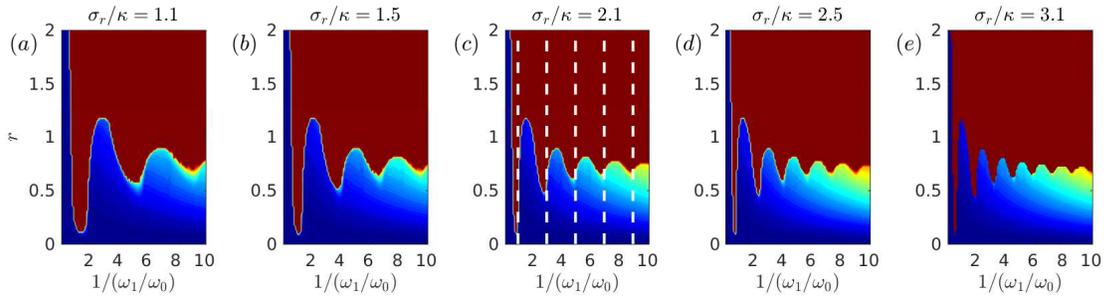}
\caption{ 
{\bf Phase diagrams for different values of $\sigma_r/\kappa$}.
(a-e) Phase diagrams for $\sigma_r/\kappa=1.1, 1.5, 2.1, 2.5, 3.1$,
respectively, where $\omega_1$ is the base frequency, i.e., the largest 
resonant frequency. The white dashed lines on (c) correspond to the cases
of $\omega_1/\omega=1, 3, 5, 7, 9$, respectively. There is a general 
relation: $\omega_{base}\propto \sigma_r/\kappa$.}
\label{fig:app}
\end{figure}

\paragraph*{Phase diagram in $1/\omega$ axis.}
As can be seen from the phase diagrams in Fig.~\ref{fig:PD}, the boundary 
between the pseudo-$\mathcal{PT}$ breaking and the pseudo-$\mathcal{PT}$ 
phases exhibits an oscillation pattern. A detailed investigation indicates
that the inverse of the peak frequencies are odd multiples of the inverse 
of the rightmost peak frequency. To better understand this behavior, 
especially in the limit $\omega\rightarrow0$, we present a series of phase 
diagrams in the $(1/\omega,r)$ plane for different values of 
$\sigma_r/\kappa$, as shown in Figs.~\ref{fig:app}(a-e), where the 
parameter values in Fig.~\ref{fig:app}(c) are the same as those in 
Fig.~\ref{fig:PD}(a). For convenience, in Fig.~\ref{fig:app}(c) the 
$1/\omega$ axis is rescaled as $1/(\omega_1/\omega_0)$, where 
$\omega_1/\omega_0=4.2$. Other panels are scaled using the base frequency 
of Fig.~\ref{fig:app}(c). In Fig.~\ref{fig:app}(c), the white dashed lines
are for $\omega/\omega_1=1,3,5,7,9$, which is indicative that, in the 
vicinity of these lines, the system is in the pseudo-$\mathcal{PT}$ 
breaking phase. As noted, we have $\omega_1/\omega_0\sim 2\sigma_r/\kappa$, 
so as $\sigma_r/\kappa$ is varied, we expect to see a similar resonance 
phenomena but with a different base frequency, as shown in 
Figs.~\ref{fig:app} (a,b,d,e). Physically, the quantity $\sigma_r/\kappa$
measures the band gap of the massive relativistic dispersion relation, and
it is known for the Hermitian case that the ZB trembling frequency is 
proportional to the band gap. As a result, the modulation frequency, which 
is at resonance with the ZB trembling frequency, is also modified. Similar 
results were obtained recently~\cite{JMDP:2014,GW:2015,LJ:2015,LHLMJL:2016,
L:2016}.

\section{Conclusion} \label{sec:conclusion}

Novel electronic materials obeying relativistic quantum mechanics are
a current focus of condensed matter physics and materials science, which
provide us with a platform to uncover, understand, and exploit unusual
physical phenomena. It is of great interest to use optical systems to
gain insights into the fundamentals of relativistic quantum solid state
devices, and vice versa, i.e., to exploit relativistic quantum electronic
behaviors to generate revolutionary ideas/methodologies in manipulating
light propagation. The main reason for such an interest is that,
electronic materials face many experimental and technological
challenges due to the extremely small wavelength issue. The conventional
method is to engineer, e.g., through strain, doping and voltage control,
some specific materials to realize the desired configuration, constraining
researchers to some limited kinds of lattice structures and interactions.
Compounded on the small wavelength issue are various kinds of disorders and
impurities. Photonic crystals and waveguides can be used to mimic
the electronic properties, e.g., the dispersion relations, of solid state
materials, with the advantage that the optical wavelength is orders of
magnitude larger than the electronic wavelength. Substantially larger
optical devices can then be fabricated to investigate a host of electronic
behaviors in relativistic quantum solid devices. These are synthetic
photonic materials and devices.

This paper is a case study of exploiting the equivalence between coupled
wave equations and the Dirac equation to model/simulate relativistic quantum 
effects in a non-Hermitian setting. Standard quantum mechanics 
requires observable operators to be Hermitian to ensure real eigenvalues. 
However, non-Hermitian Hamiltonians are also capable of generating a
purely real eigenvalue spectrum insofar as the system has a $\mathcal{PT}$
symmetry. In optics, non-Hermitian Hamiltonian systems can be
realized using materials with a complex index of refraction, rendering
experimentally feasible such systems~\cite{Rueter:2010,Regensburger:2012,
Peng:2014,Wimmer:2015}. We studied a class of non-Hermitian waveguide systems
with periodic modulated imaginary refraction index in order to uncover the
optical counterpart of the relativistic ZB effect, which exhibit
$\mathcal{PT}$ symmetry breaking at any instant of time but the symmetry is
preserved on a larger time scale (herewith the term {\em pseudo $\mathcal{PT}$
symmetry}). We generated phase diagrams of the pseudo-$\mathcal{PT}$ behavior,
which provide a general picture for controlling and harnessing light propagation
in the waveguide system. The phenomenon of oscillatory boundary variations
in the phase diagrams is explained. In the low modulation frequency region, we
observed periodically enhanced and suppressed ZB oscillations, as predicted by
an analysis of the equivalent Dirac equation. The phenomenon that there are time
periods where the ZB trembling is absent, interspersed by time intervals where
the relativistic quantum oscillations emerge and disappear, a kind of
``revival'' phenomenon, can potentially lead to a new mechanism to
manipulate/control light propagation. In the high frequency regime, a 
conventional ZB trembling behavior arises due to a mean-field effect. In the
intermediate region, the equivalent description based on the Dirac equation no
longer holds. However, numerically we uncovered a spatial energy localization
phenomenon.

\section*{Acknowledgement}

This work was supported by AFOSR under Grant No.~FA9550-15-1-0151 and
by ONR under Grant No.~N00014-16-1-2828. L.H. was also supported by NSF 
of China under Grant No.~11375074.

\appendix

\paragraph*{Derivation of the main analytical result Eq.~(\ref{eq:ana}).}
We start from the Dirac equation in the momentum space:
\begin{equation*}
i\hbar\frac{\partial\psi(k,t)}{\partial t} =
\hbar[\kappa k\alpha_1+\sigma(t)\alpha_3]\psi(k,t).
\end{equation*}
The wavefunction at any given time can be written as
\begin{equation*}
\psi_k(t)=\mathcal{T}e^{-\frac{i}{\hbar}\int_0^tH(t')dt'}\psi_k(0).
\end{equation*}
where $\mathcal{T}$ is the time ordering operator. Since our system is a
time dependent non-Hermitian system, i.e., $[H(t_1),H(t_2)]\neq0$, the
expansion of the exponential factor under the time ordering operator can be
quite sophisticated. In order to gain analytical insights, we assume
$[H(t_1),H(t_2)]=0$. This does not mean that the non-Hermitian property of 
our system is lost. In fact, non-Hermitian characteristics still exist 
in $E_{\pm}(t)$.  

We first discuss the validity of the approximations. The effective 
Hamiltonian of the system is $H(k,t)=\kappa k\alpha_1+\sigma(t)\alpha_3$,
where time modulation occurs in the imaginary part of $\sigma$: 
$\sigma(t)=\sigma_r+i\sigma_isin(\omega t)$, while the real part 
$\sigma_r$ is time independent. In the limit $\sigma_i=0$, system 
becomes time independent, rendering irrelevant the time ordering operator, as 
$lim_{\sigma_i\rightarrow0}[H(t_1),H(t_2)]\propto\kappa k\sigma_i\rightarrow0$.
The time dependent weight on the $\alpha_1$ term comes from the integral 
$\int_0^tH(t')dt'$, which is responsible for the dynamical phase 
in the limit.

For a small value of $\sigma_i$, the time ordering operator can be 
written as
\begin{eqnarray*}
    \mathcal{T}[e^{-i\int_0^tH(t')dt'}]=I&+\frac{-i}{1!}\int_0^tdt_1\mathcal{T}[H(t_1)] \\
    &+\frac{(-i)^2}{2!}\int_0^tdt_1\int_0^tdt_2\mathcal{T}[H(t_1)H(t_2)] \\
    &+\dots \\
    &+\frac{(-i)^k}{k!}\int_0^tdt_1\dots\int_0^tdt_k\mathcal{T}[H(t_1)\dots H(t_k)] \\
    &+\dots,
\end{eqnarray*}
where, for simplicity, we have ignored $\hbar$. Take the second order term
$\int_0^tdt_1\int_0^tdt_2 \mathcal{T}[H(t_1)H(t_2)]$ as an example. The
difference induced by the time ordering operator is on the order of
$[H(t_1),H(t_2)]\propto \kappa k\sigma_i$, while the main part of the
second order term is dominated by $\kappa^2k^2+\sigma_r^2$. Since we
focus on the excitations about the $k\sim0$ points, i.e., the Brillouin
zone boundary in the original waveguide system, the ratio of the two
terms, $\kappa k\sigma_i/(\kappa^2k^2+\sigma_r^2)$, is negligibly
small, providing a justification that the effect of the time ordering
operator can be neglected in the regime
$\sigma_i/\sigma_r\rightarrow0$. For
higher order terms, the effect of the time ordering operator will be
even more negligibly small.

Taken together, for $\sigma_r \ne 0$, our results are exact for 
$\sigma_i=0$, as the system reduces to a time independent system.
For $\sigma_i << \sigma_r$ (e.g., $\sigma_i/\sigma_r=0.2$), the effects of the
time ordering operator can be neglected. From the comparison with the numerical
results, we can see the approximation works well when $\sigma_i/\sigma_r=0.2$.
This agreement demonstrates that the oscillations we observed are indeed
Zitterbewegung, which possesses a modulated amplitude resulting from the
non-Hermitian modulation of the system.
When $\sigma_i/\sigma_r$ increases to $0.5$, our analytical results cannot
quantitatively reproduce the numerical results. However, the positions which are
of tremendous oscillations are well predicted.

Since $\sigma(t)=\sigma_r+i\sigma_i\sin(\omega t)$, we have
\begin{eqnarray}
\nonumber 
-\frac{1}{\hbar}\int_0^tH(t')dt' & = & -\kappa kt\alpha_1
+(-\sigma_rt+i\sigma_i\frac{\cos(\omega t)-1}{\omega})\alpha_3 \\ \nonumber
& \equiv & A(\hat{A}\cdot\overrightarrow{\alpha}), 
\end{eqnarray}
where 
\begin{eqnarray}
\nonumber
\overrightarrow{A} & = & A\cdot\hat{A}=(-\kappa kt,0,-\sigma_rt
+i\sigma_i\frac{\cos(\omega t)-1}{\omega}), \\ \nonumber 
A & = & \sqrt{\kappa^2k^2t^2+(\sigma_rt-i\sigma_i
\frac{\cos(\omega t)-1}{\omega})^2}, \ \ \mbox{and} \\ \nonumber 
\overrightarrow{\alpha} & = & (\alpha_1,\alpha_2,\alpha_3)^T.
\end{eqnarray}
Using the Euler formula for Pauli matrices,
\begin{equation}
e^{iA(\hat{A}\cdot\overrightarrow{\alpha})}
=\hat{I}\cos(A)+i(\hat{A}\cdot\overrightarrow{\alpha})\sin(A),
\label{}
\end{equation}
we have
\begin{equation}
    \psi_k(t) \approx e^{-\frac{i}{\hbar}\int_0^tH(t')dt'}\psi_k(0) \\ \nonumber
    =[I\cos(A)+i(\hat{A}\cdot\overrightarrow{\alpha})\sin(A)]\psi_k(0).
\end{equation}
Next we consider the initial condition. At the input
plane $z=0$, the initial wave packet can be chosen to have the shape of a
slowly varying Gaussian form. The initial amplitude of $\psi_{1,2}(\xi,0)$ 
is then proportional to $G(2na)$ and $G((2n-1)a)\approx G(2na)$, so
the two components of the Dirac spinor have the same initial condition,
i.e., $\psi_k(0)=G(k)[1,1]^T$. The Pauli matrices operate on the
initial Dirac spinor, and we get the time evolution of the two
components of the Dirac spinor as
\begin{equation}
    \psi_k(t)=G(k) \left( \begin{array}{cc}
    \cos(A)+i\frac{\sin(A)}{A}[-\kappa kt+(-\sigma_rt+i\frac{\cos(\omega t)-1}{\omega}\sigma_i)] \\
    \cos(A)+i\frac{\sin(A)}{A}[-\kappa kt-(-\sigma_rt+i\frac{\cos(\omega t)-1}{\omega}\sigma_i)]
                          \end{array} \right).
\end{equation}
The expectation value of the position operator can be calculated through
\begin{equation}
    \langle\xi\rangle(t)=2\pi
    i\int_{-\infty}^\infty\psi_k^*(t)\partial_k\psi_k(t)
\end{equation}

A lengthy algebra leads to the main analytic result: the formulas in
Eq.~(\ref{eq:ana}) as well as the time evolution of the modulus of the 
wavefunction. The convergence of the integrals in Eq.~(\ref{eq:ana}) is
guaranteed by the exponential decay of the Gaussian wave spectrum
$G(k)$. In particular, for small $t$ values, $|\cos(A)|^2$ is
about unity. For large values of $t$, we have
$|\psi(t)|^2\approx|\cos(A)|^2+|\sin(A)|^2\geqslant1$. Thus, $|\psi(t)|^2$
can never approach $0$, so Eq.~(6) converges.
 
\paragraph*{Derivation of the dispersion relation.}
We follow Refs.~\cite{SK:2002,L:2010} to derive the dispersion relation. 
Specifically, from the coupled mode equation, we obtain, for $a_{2n}$
and $a_{2n+1}$, the following equations:
\begin{eqnarray}
\nonumber
i\frac{da_{2n}}{dz} & = & -\kappa(a_{2n-1}+a_{2n+1})+\sigma(z)a_{2n}, \\ \nonumber
i\frac{da_{2n+1}}{dz}& = & -\kappa(a_{2n}+a_{2n+2})-\sigma(z)a_{2n+1},
\end{eqnarray}
where $\sigma(z)$ is in general complex. As there are two sets of
waveguides, we assume $a_{2n}\sim Ae^{iq(2n)a-i\omega z}$ and
$a_{2n+1}\sim Be^{iq(2n+1)a-i\omega z}$. Substituting these into the
equations for $a_{2n}$ and $a_{2n+1}$, after some algebra, we have
\begin{eqnarray}
        (\omega-\sigma(z))A+2\kappa \cos(qa)B & = & 0, \\ \nonumber
        2\kappa \cos(qa)A+(\omega+\sigma(z))B & = & 0.
\end{eqnarray}
The determinant of the above equation needs to be $0$, which leads to
$\omega=\pm\sqrt{\sigma(z)^2+4\kappa^2\cos^2(qa)}$. To account for the fact
that $\sigma(z)$ is complex, we substitute $\sigma(z)=\sigma_r+i\sigma_i(t)$
into the dispersion relation and consider the limit of small imaginary part:
\begin{eqnarray}
\omega &=&\pm\sqrt{(\sigma_r+i\sigma_i(z))^2+4\kappa^2\cos^2(qa)} \\ \nonumber
&=&\pm\sqrt{\sigma_r^2-\sigma_i^2(z)+4\kappa^2\cos^2(qa)}\cdot 
\sqrt{1+\frac{2i\sigma_r\sigma_i}{\sigma_r^2-\sigma_i^2(z)+4\kappa^2\cos^2(qa)}} 
\\ \nonumber
& \approx&\pm\sqrt{\sigma_r^2-\sigma_i^2(z)+4\kappa^2\cos^2(qa)}\cdot 
(1+\frac{i\sigma_r\sigma_i}{\sigma_r^2-\sigma_i^2(z)+4\kappa^2\cos^2(qa)}).
\end{eqnarray}
The real part of the dispersion relation in the limit of small imaginary
part is then given by
\begin{equation}
\pm\sqrt{\sigma_r^2-\sigma_i^2(z)+4\kappa^2]\cos^2(qa)}.
\end{equation}


\providecommand{\newblock}{}

\end{document}